\documentclass[conference]{IEEEtran}

\usepackage{cite}
\usepackage{amsmath,amssymb,amsfonts}
\usepackage{algorithmic}
\usepackage{graphicx}
\usepackage{textcomp}
\usepackage{xcolor}
\usepackage{hyperref}
\usepackage{booktabs}
\usepackage{xurl}
\usepackage{multirow}
\usepackage{placeins}
\usepackage{tikz}
\usepackage{pgfplots}
\usetikzlibrary{arrows.meta, positioning, calc, fit, backgrounds, shapes.geometric}
\usepgfplotslibrary{groupplots}
\pgfplotsset{compat=1.18}
\definecolor{cLeader}{HTML}{C44E52}   
\definecolor{cRead}{HTML}{4C72B0}     
\definecolor{cNode}{HTML}{55A868}     
\definecolor{cMuted}{HTML}{8C8C8C}    
\definecolor{cClient}{HTML}{4A4A4A}   

\tikzset{
  member/.style   = {rectangle, rounded corners=2pt, draw=cNode!80!black,
                     fill=cNode!12, thick, minimum width=15mm, minimum height=8mm,
                     font=\footnotesize\sffamily, align=center},
  leader/.style   = {member, draw=cLeader!80!black, fill=cLeader!12},
  client/.style   = {rectangle, rounded corners=2pt, draw=cClient, fill=cClient!8,
                     thick, minimum width=13mm, minimum height=7mm,
                     font=\footnotesize\sffamily, align=center},
  db/.style       = {cylinder, shape border rotate=90, aspect=0.25, draw=cNode!80!black,
                     fill=cNode!8, minimum width=9mm, minimum height=6mm,
                     font=\scriptsize\sffamily, align=center},
  wflow/.style    = {-{Latex[length=2.4mm]}, cLeader, line width=1.6pt},   
  rflow/.style    = {-{Latex[length=2mm]}, cRead, line width=0.8pt},       
  repl/.style     = {-{Latex[length=1.8mm]}, cLeader!70, densely dashed, line width=0.9pt},
  cflow/.style    = {-{Latex[length=2mm]}, cClient, line width=1pt},       
  serve/.style    = {-{Latex[length=2mm]}, cRead, line width=1pt},
  elabel/.style   = {font=\scriptsize\sffamily, inner sep=1.5pt, align=center},
  note/.style     = {font=\scriptsize\itshape, text=cMuted, align=center},
  panel/.style    = {font=\small\sffamily\bfseries},
}

\begin{document}

\title{Operation-Type-Aware Client Routing for Leader-Based Consensus Datastores}

\author{
\IEEEauthorblockN{Sri Saran Balaji Vellore Rajakumar}
\IEEEauthorblockA{\textit{Amazon Web Services} \\
Seattle, WA, USA \\
srajakum@amazon.com}
\and
\IEEEauthorblockN{James Thompson}
\IEEEauthorblockA{\textit{Amazon Web Services} \\
Seattle, WA, USA \\
jamesmt@amazon.com}
\and
\IEEEauthorblockN{Gyuho Lee}
\IEEEauthorblockA{\textit{NVIDIA} \\
gyuhol@nvidia.com}
}

\maketitle

\begin{abstract}
Leader-based consensus datastores (etcd, ZooKeeper) face two competing routing goals: spread load evenly across members, and route operations to the member whose protocol role matches the operation. Writes must commit through the leader, so sending them elsewhere adds a forwarding hop. Linearizable reads need only a lightweight leader confirmation before any member can serve them locally. The upstream etcd client uses gRPC's \texttt{round\_robin} balancer, distributing reads and writes uniformly across cluster members. An operation-aware client resolves this by pinning writes to the leader and distributing reads across the healthy read pool. In steady state on a 3-node etcd cluster (80/20 read/write mix, 5~trials), this lowers write P50 by 29\% and raises throughput by 9\%. When a follower degrades silently, the operation-aware client detects the latency shift and removes it from the read pool, cutting read P99 by 64\%, write P99 by 74\%, and raising throughput by 89\%. The same routing rule applied to ZooKeeper (ZAB protocol, different implementation) points in the same direction, suggesting the result follows from leader-based consensus structure rather than one system's implementation. The key obstacle to discovering this policy adaptively is that the leader confirmation round-trip occurs between cluster members, so the client sees only a blended latency signal rather than the decisive coordination cost directly.
\end{abstract}

\begin{IEEEkeywords}
etcd, consensus systems, adaptive routing, Kubernetes, gRPC load balancing, linearizable reads, availability zones
\end{IEEEkeywords}

\section{Introduction}
\label{sec:introduction}

Leader-based strongly-consistent datastores such as etcd~\cite{etcdraft} and Apache ZooKeeper~\cite{hunt2010zookeeper} are foundational components in many infrastructure systems. Both implement the replicated state machine model~\cite{schneider1990statemachine} over consensus protocols, Raft~\cite{etcdraft} and Zab~\cite{junqueira2011zab} respectively. Many production systems place frequent coordination traffic on etcd and ZooKeeper. ZooKeeper serves Kafka~\cite{kreps2011kafka} and Pulsar, among others. In these settings the datastore becomes a synchronous bottleneck. When it saturates, the application stalls. Even a few milliseconds per operation matter when the datastore lies on the critical path. These datastores run as clusters whose members cooperate to provide durability and availability, but the members are not interchangeable. At any point one member acts as the leader and the others follow it, so each node carries a different protocol responsibility. Yet production clients still tend to apply a stateless load-balancing assumption and treat cluster members as interchangeable endpoints. A stock etcd client routes every request, read and write alike, through gRPC's \texttt{round\_robin} balancer, and a ZooKeeper client similarly chooses one server from its connect string and stays with it until reconnect, again without regard to protocol role. That default is only partly correct because healthy reads can still benefit from distribution, but writes must commit through the leader and gain nothing from being sent to a follower first. Once a member becomes a bad read target, continuing to give it an equal share of reads is also inefficient.

Our claim is that client routing in leader-based consensus should follow operation semantics rather than one destination policy or client-observed latency alone, especially in latency-sensitive systems where the datastore lies on the synchronous critical path. The main etcd tables in this paper are steady-state measurements taken over intervals in which the leader remained fixed, so we separate the steady-state claim from the failover claim. In the healthy steady state, linearizable reads should stay distributed across the members that can serve them, while writes should go to the leader. A practical client still needs active leader refresh and fallback when an election occurs, along with a read-side monitor that stops sending traffic to a member that has become a clear latency outlier. This paper supplies a measured decomposition: why healthy read distribution is right, why writes should not follow the same rule, where a latency signal still helps, and where the rule stops holding. In our data, the strongest practical benefit appears when a nearby follower becomes slow. A client that keeps reading from it blindly pays a large tail penalty, while an operation-aware client can preserve leader-pinned writes and almost entirely remove that endpoint from the read pool. The same split also improves the healthy write path, and the repeated-trial results show that those write gains remain material on both 3-node and 5-node clusters.

\subsection{The Operation-Type Asymmetry}

A write commits through the leader. A follower that receives a write forwards the entire proposal to the leader, which alone drives it through consensus, so sending a write anywhere but the leader adds a forwarding hop and nothing else. With \texttt{round\_robin} on a three-node cluster, $2/3$ of writes pay this hop. On a five-node cluster, $4/5$ do. Routing writes to the leader removes it.

A linearizable read places the opposite demand on the leader, which must confirm the latest commit index but need not serve the read itself. etcd's default linearizable read uses the ReadIndex protocol. The serving member asks the leader for a commit index confirmation, waits until its own applied state reaches that point, and then serves the response from its own local storage. The leader coordinates the read, but it does not receive the query payload or assemble the reply. ZooKeeper's \texttt{sync()} plays the same role, and that shared structure is the core observation behind the paper. Followers can serve linearizable reads~\cite{herlihy1990linearizability}, so distributing reads across $M$ members lets the client use the read-serving resources of the full cluster while preserving linearizability.

These two facts point in opposite directions. The stock client gets the read side roughly right in the healthy case because it spreads read-serving work across the cluster. It gets the write side wrong because most writes still land on followers and pay forwarding, and it keeps getting the read side wrong once one member becomes a bad read target. Routing everything to the leader concentrates read-serving work on one node, and we measure read latency rising by up to 178\% under pure-read load as the leader's storage engine becomes the bottleneck. Distributing everything spends write bandwidth on forwarding and keeps hitting a slow reader during a grey failure. The right policy is therefore not another heuristic for picking one endpoint but a decomposition by operation type, with writes going to the leader and reads remaining distributed unless a member becomes clearly degraded. This keeps the leader, the one component that cannot be scaled out, from becoming an unnecessary read hotspot while still giving the client a way to isolate bad read endpoints.

\subsection{Consensus-Cost Invisibility}

\begin{figure}[t]
  \centering
  \begin{tikzpicture}
    \node[client] (bc) at (0,0) {Latency-adaptive\\client};

    \node[member] (ef) at (3.6,1.2) {Follower\\(near)};
    \node[leader] (el) at (3.6,-1.2) {\textbf{Leader}\\(far)};

    \draw[cflow, line width=2.0pt, cRead] (bc) -- (ef);
    \draw[cflow, line width=1.1pt, cLeader] (bc) -- (ef);
    \node[elabel, text=cRead] at ($(bc)!0.45!(ef) + (0,0.3)$) {many reads};
    \node[elabel, text=cLeader] at ($(bc)!0.55!(ef) + (0,-0.35)$) {few writes};
    \draw[cflow, cMuted] (bc) -- (el);
    \node[elabel, text=cMuted] at (2.0,-1.5) {higher RTT};
    \node[note, text width=34mm, text=cRead] at (1.55,2.15)
      {Observed latency is dominated by the many fast reads to the nearby follower.};

    \draw[cMuted, line width=1pt, dash pattern=on 3pt off 2pt] (-1.3,-2.55) -- (5.8,-2.55);
    \node[fill=white, inner sep=1.5pt, font=\scriptsize\sffamily\bfseries, text=cMuted]
      at (2.2,-2.55) {\ \ visibility boundary\ \ };
    \node[font=\scriptsize\sffamily, text=cClient] at (-0.05,-2.2) {observed};
    \node[font=\scriptsize\sffamily, text=cMuted]  at (-0.05,-2.9) {hidden};

    \node[member, draw=cMuted, fill=cMuted!8, text=cMuted] (hf) at (1.2,-3.75) {near\\follower};
    \node[leader, draw=cMuted, fill=cMuted!8, text=cMuted] (hl) at (4.6,-3.75) {leader};

    \draw[rflow, cMuted] (hf.north east) to[bend left=15]
        node[elabel,above=2pt,text=cMuted]{light read confirm} (hl.north west);
    \draw[wflow, cMuted] (hf.east) -- node[elabel,above=2pt,text=cMuted]{write forwarding} (hl.west);

    \node[note, text width=55mm] at (2.8,-4.7)
      {The client sees one blended endpoint latency. The fewer writes sent to
       the nearby follower still pay hidden follower-to-leader work, but they
       are diluted by the much larger volume of fast reads.};
  \end{tikzpicture}
  \caption{Consensus-cost invisibility. A latency-adaptive client sees one
  blended latency signal per endpoint. Under a mixed workload, many fast reads
  to a nearby follower can make that follower look best overall even though the
  fewer writes sent there still pay hidden follower-to-leader forwarding. That
  is why latency alone does not recover the correct in-region steady-state
  routing rule.}
  \label{fig:invisibility}
\end{figure}
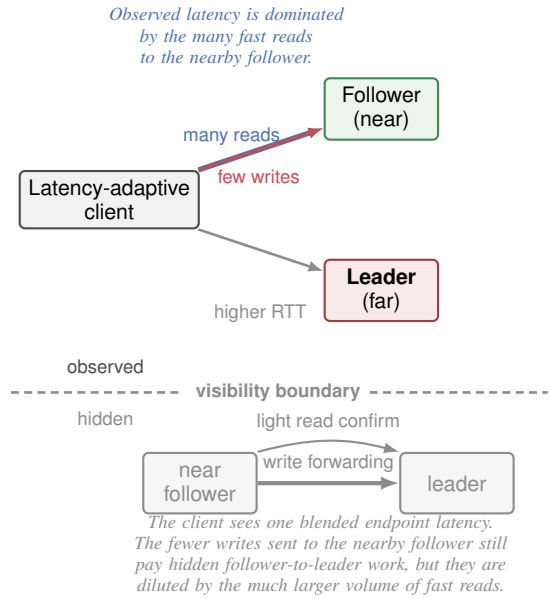

A natural question is why an adaptive client, one that simply learns the fastest endpoint, does not discover this on its own. The reason is that the signal it would need is hidden, which we summarize as the following principle.

\begin{quote}
\emph{Consensus-cost invisibility.} In a leader-based consensus system, the coordination cost an operation triggers inside the cluster is paid mostly server-side and is not cleanly exposed to the client. Client-observed latency is a blended end-to-end signal. Under a mixed workload, many fast reads to a nearby follower can dominate that signal even while the fewer writes sent there still pay follower-to-leader forwarding inside the cluster. Latency alone therefore does not reliably identify the correct steady-state endpoint.
\end{quote}

This is why latency-based load balancing, the usual choice for stateless backends, is a poor fit here (Figure~\ref{fig:invisibility}). In our in-region experiments it does not recover the correct steady-state endpoint class. A nearby follower can still look best to the client because its large volume of reads dominates the observed latency signal, even though the writes sent there still pay hidden forwarding through the leader. Routing is better driven by operation semantics, which the application already knows, than by observed latency alone. The same latency signal still matters, but in a narrower role. It is useful for detecting degraded members that should leave the read pool.

\subsection{Contributions}

\begin{enumerate}
    \item \textbf{A comparison of stock \texttt{round\_robin} and \texttt{hybrid}, with latency-adaptive Boltzmann as a diagnostic comparator.} The main comparison is \texttt{round\_robin} vs.\ \texttt{hybrid} under a stable leader; Boltzmann serves to show where latency adaptation helps and where it does not.
    \item \textbf{A measured decomposition by operation type.} Eliminating write forwarding lowers write P50 by 29--37\% (3 and 5~nodes). Distributed reads outperform leader-concentrated reads by up to 178\%. Under grey failure (8~ms delay on a same-AZ follower), hybrid cuts read P99 by 64\% and write P99 by 74\%.
    \item \textbf{The principle of consensus-cost invisibility.} Client-observed latency does not recover the correct steady-state split, but remains useful for degradation detection.
    \item \textbf{Cross-system validation on ZooKeeper (ZAB),} suggesting the result reflects leader-based consensus rather than one implementation.
\end{enumerate}

\section{Background}
\label{sec:background}

\begin{figure*}[t]
  \centering
  \begin{tikzpicture}
    \begin{scope}[local bounding box=W]
      \node[panel] at (2.1,3.0) {(a) Write: forwarding adds work and nothing else};

      \node[client] (wc)  at (0,1.4) {Client};
      \node[member] (wf)  at (2.4,1.4) {Follower};
      \node[leader] (wl)  at (2.4,-0.6) {\textbf{Leader}};
      \node[member] (wf2) at (4.8,1.4) {Follower};

      \draw[cflow] (wc) -- node[elabel,above]{write} (wf);
      \draw[wflow] (wf) -- node[elabel,right=1pt,text=cLeader]{forward\\proposal} (wl);
      \draw[repl] (wl.east) to[bend right=18]
            node[elabel,below right=-1pt and -1pt]{replicate} (wf2.south);
      \draw[repl] (wl.north) .. controls +(-0.9,0.5) and +(-0.5,-0.7) .. (wf.south);

      \node[note, text width=46mm] at (2.4,-2.15)
        {A follower cannot commit the write. It forwards the proposal to the
         leader, which alone drives it through consensus.\\
         With \texttt{round\_robin}, $(M{-}1)/M$ of writes pay this extra hop.};
    \end{scope}

    \draw[cMuted!50, line width=0.6pt] (5.8,-2.4) -- (5.8,3.2);

    \begin{scope}[shift={(7.0,0)}, local bounding box=R]
      \node[panel] at (2.8,3.0) {(b) Read: lightweight confirmation, local serve};

      \node[client] (rc)  at (0,1.4) {Client};
      \node[member] (rf)  at (2.6,1.4) {Serving\\member};
      \node[leader] (rl)  at (2.6,-0.6) {\textbf{Leader}};
      \node[db]     (rdb) at (5.2,1.4) {local\\state};

      \draw[cflow] (rc) -- node[elabel,above=8pt]{read} (rf);
      \draw[rflow] (rf.south) to[bend right=12]
            node[elabel,left=1pt,text=cRead]{lightweight\\confirm} (rl.north);
      \draw[rflow, densely dotted] (rl.north east) to[bend right=12]
            node[elabel,right=1pt,text=cRead]{commit idx} (rf.south east);
      \draw[serve] (rf) -- node[elabel,above]{local serve} (rdb);

      \node[note, text width=48mm] at (2.4,-2.15)
        {The leader only confirms that the serving member is up to date enough
         to answer. The query payload is not passed to the leader.\\
         Healthy members can therefore share the read-serving work.};
    \end{scope}
  \end{tikzpicture}
  \caption{The operation-type asymmetry. A write (a) must be driven through
  the leader, so routing it to a follower adds a forwarding hop and nothing
  else. A linearizable read (b) only needs a lightweight confirmation from the
  leader that the receiving member is current, after which that member serves
  the read from local state. The two operation types therefore want opposite
  routing rules: writes to the leader, reads distributed across healthy serving
  members.}
  \label{fig:asymmetry}
\end{figure*}

The operation-type asymmetry from Section~\ref{sec:introduction} is not an implementation detail of one system. It follows from how leader-based consensus serves the two operation types (Figure~\ref{fig:asymmetry}). We describe the mechanism first in etcd (Raft) and then show that ZooKeeper (ZAB) provides the same structure, which is why the same routing rule applies to both.

\subsection{Writes Are Leader-Only, So Forwarding Is Pure Overhead}

In both etcd and ZooKeeper, a write is committed by replicating it through the leader to a quorum, the standard structure of a replicated state machine~\cite{schneider1990statemachine} built on a leader-based consensus protocol~\cite{etcdraft, junqueira2011zab}. A follower that receives a client write does not commit it locally. It \emph{forwards} the entire proposal to the leader and relays the result back. Sending a write to a follower therefore provides no benefit and adds a network hop. With a round-robin client on an $M$-node cluster, a fraction $(M-1)/M$ of writes pay this forwarding penalty, $2/3$ at three nodes and $4/5$ at five. Routing writes directly to the leader removes that extra hop.

\subsection{Linearizable Reads Need Leader Confirmation but Followers Serve}

The reason reads behave oppositely is that a linearizable read does not require the leader to \emph{serve} it, only to \emph{confirm} the follower is current. etcd implements this with the ReadIndex protocol~\cite{etcdraft}. The follower sends a \texttt{MsgReadIndex} to the leader, receives a commit index confirmation, waits until its own applied index reaches that point, and then serves the response from local storage. The query payload is not passed to the leader, and the leader does not build the response. For routing, the important point is that any follower can serve a linearizable read after a lightweight coordination step, so distributing reads across $M$ members can use the read-serving resources of all $M$ members (CPU, storage engine, bandwidth) while preserving linearizability. At high concurrency a single node's storage engine becomes CPU-bound, so concentrating reads on the leader wastes the other members' serving capacity.

\subsection{The Same Mechanism in ZooKeeper (ZAB)}
\label{sec:bg-zk}

ZooKeeper's ZAB protocol has the same structure. Writes are totally ordered through the leader, and a follower forwards client writes to it, paying the same forwarding hop. For reads, ZooKeeper's default read is served locally by whatever server the client is connected to and may be slightly stale. A client that requires a linearizable read issues \texttt{sync()} before the read, which forces the connected follower to catch up to the leader's latest committed transaction before serving. \texttt{sync()} is thus the direct analogue of etcd's ReadIndex. It is a lightweight leader-coordination step after which a follower serves the read from its own copy. Because the mechanism is the same, the routing rule, with writes to the leader and linearizable reads distributed across all members, carries over as well. So does its justification in the operation-type asymmetry, which we confirm in Section~\ref{sec:eval-zookeeper}.

\subsection{Client Load Balancing}

The etcd v3 client uses gRPC's \texttt{round\_robin} balancer via the client library's internal resolver~\cite{grpc}. ZooKeeper's client picks one server from its connect string and stays with it. General-purpose service proxies such as Envoy~\cite{envoy} offer richer balancing, but remain oblivious to consensus roles. Neither is operation-type- or consensus-aware. Both treat members as interchangeable, which is exactly the assumption the asymmetry violates.

\subsection{HTTP/2 Head-of-Line Blocking}

gRPC multiplexes RPCs over a single HTTP/2 connection per endpoint, so when many requests concentrate on one endpoint, stream scheduling contention on that connection can become the bottleneck. Connection pooling (multiple connections per endpoint) relieves this, but only when per-connection head-of-line blocking actually dominates. We therefore hold connection count equal across policies for fairness, not as a universal requirement.

\tikzset{
  mnode/.style = {circle, draw=cNode!80!black, fill=cNode!12, thick,
                  minimum size=6mm, font=\scriptsize\sffamily, inner sep=0pt},
  mlead/.style = {mnode, draw=cLeader!80!black, fill=cLeader!15},
  meject/.style= {mnode, draw=cMuted, fill=cMuted!10, text=cMuted, densely dashed},
  mcli/.style  = {rectangle, rounded corners=1pt, draw=cClient, fill=cClient!8,
                  font=\scriptsize\sffamily, inner sep=1.5pt, minimum height=5mm},
  wr/.style    = {-{Latex[length=1.6mm]}, cLeader, line width=1.1pt},
  rd/.style    = {-{Latex[length=1.4mm]}, cRead, line width=0.7pt},
  rdw/.style   = {-{Latex[length=1.6mm]}, cRead, line width=1.3pt},
  rdt/.style   = {-{Latex[length=1.2mm]}, cRead, line width=0.4pt},
  ptitle/.style= {font=\scriptsize\sffamily\bfseries, align=center},
}
\newcommand{\minicluster}[2]{%
  \begin{scope}[shift={(#2,0)}]
    \node[mcli] (#1-c)  at (0,0)        {C};
    \node[mlead] (#1-l) at (1.55,0.72)  {L};
    \node[mnode] (#1-f1) at (1.55,0)    {F};
    \node[mnode] (#1-f2) at (1.55,-0.72){F};
  \end{scope}
}
\begin{figure*}[t]
  \centering
  \begin{tikzpicture}[node distance=0pt]
    \def\dx{4.9}

    \minicluster{rr}{0}
    \draw[rd] (rr-c) -- (rr-l);
    \draw[rd] (rr-c) -- (rr-f1);
    \draw[rd] (rr-c) -- (rr-f2);
    \draw[wr] (rr-c.north) to[bend left=18] (rr-l.west);
    \draw[wr] (rr-c.south) to[bend right=14] (rr-f1.west);
    \draw[wr] (rr-c.south) to[bend right=4] (rr-f2.west);
    \node[ptitle] at (0.8,-1.65) {\texttt{round\_robin}\\{\scriptsize same rule for both operations}};

    \minicluster{bz}{\dx}
    \draw[rdw] (bz-c) -- (bz-f1);
    \draw[rd]  (bz-c) -- (bz-f2);
    \draw[rdt] (bz-c) -- (bz-l);
    \draw[wr]  (bz-c.north) to[bend left=18] (bz-f1.west);
    \draw[wr]  (bz-c.south) to[bend right=8] (bz-f2.west);
    \node[ptitle] at ({\dx+0.8},-1.65) {\texttt{boltzmann}\\{\scriptsize one blended latency signal}};

    \minicluster{hy}{2*\dx}
    \draw[rd] (hy-c) -- (hy-l);
    \draw[rd] (hy-c) -- (hy-f1);
    \draw[rd] (hy-c) -- (hy-f2);
    \draw[wr] (hy-c.north) to[bend left=18] (hy-l.west);
    \node[meject, scale=0.85] at ({2*\dx+3.1},0.98) {F};
    \draw[rd, densely dashed] ({2*\dx+2.8},0.6) -- ({2*\dx+3.0},0.88);
    \node[elabel, anchor=west, text=cMuted] at ({2*\dx+3.25},0.98) {remove only if slow};
    \node[ptitle] at ({2*\dx+0.8},-1.65) {\texttt{hybrid}\\{\scriptsize writes to leader, healthy read pool}};

    \draw[wr] (3.9,1.55) -- ++(0.7,0);
    \node[elabel,right] at (4.6,1.55) {write};
    \draw[rd] (6.9,1.55) -- ++(0.7,0);
    \node[elabel,right] at (7.6,1.55) {read};
    \node[meject, scale=0.7] at (9.95,1.55) {};
    \node[elabel,right] at (10.3,1.55) {degraded read outlier};
  \end{tikzpicture}
  \caption{The three routing policies. Writes (red) and linearizable reads
  (blue) from a client \textbf{C} to a cluster of one leader \textbf{L} and
  two followers \textbf{F}. \texttt{round\_robin} applies one destination rule
  to every operation. \texttt{boltzmann} uses one blended latency signal for
  both reads and writes. \texttt{hybrid} splits by protocol role: writes go to
  the leader, while reads stay distributed across the healthy read pool and only use
  latency to remove a clearly degraded read endpoint.}
  \label{fig:policies}
\end{figure*}
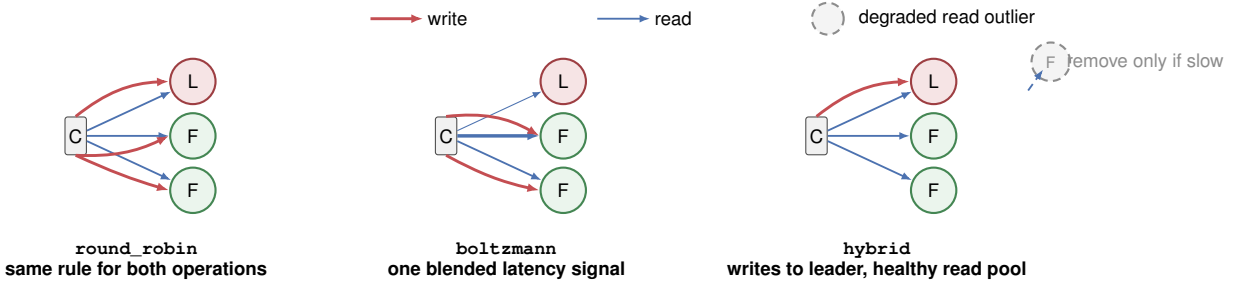

\newpage

\section{Routing Policies}
\label{sec:design}

To test the operation-type asymmetry against the alternatives, we implement three routing policies and compare them under identical conditions (Figure~\ref{fig:policies}). \texttt{round\_robin} is the stock policy and spreads every operation. \texttt{boltzmann} is the natural adaptive baseline and follows observed latency. \texttt{hybrid} is our proposed policy. It sends writes to the leader and keeps reads distributed across the healthy read pool. The point of the comparison is not to advocate another generic balancer, but to show that endpoint selection in leader-based consensus has to split by protocol role.

We implement all policies in a custom Go benchmark client that wraps the etcd v3 client library. The client maintains a pool of gRPC connections to each etcd member (default 3 per endpoint, 9 total) and selects which connection to use per request based on the active routing policy.

\subsection{Request Data Path}

Each benchmark worker generates a read or write on a random key, asks the active routing policy to choose a target endpoint, acquires one connection from that endpoint's pool, and then executes the request through the etcd v3 client library. When the request completes, the worker records the latency and, for adaptive policies, feeds that observation back into the router.

Connection pooling can matter when many concurrent workers target a single endpoint, since a single HTTP/2 connection may create stream scheduling contention (head-of-line blocking). Regardless, we allocate equal total connection counts across all policies for fair comparison, so that pooling is never a confound between policies.

\subsection{Round Robin (Baseline)}
\label{sec:design-rr}

gRPC's built-in \texttt{round\_robin} balancer distributes requests uniformly across all endpoints. Each shared client holds connections to all $M$ members. The gRPC resolver cycles through them. We create $N$ shared clients, giving $N$ total connections to the cluster. This is the stock etcd client behavior, and it applies the same destination to reads and writes alike.

\subsection{Boltzmann Exploration}
\label{sec:design-boltzmann}

Boltzmann is an adaptive policy that favors endpoints with lower recent latency but still keeps some probability of exploring the others. Each endpoint maintains an exponential weighted moving average (EWMA) of observed request latency. The router converts those latency estimates into selection probabilities with a softmax controlled by a temperature $\tau$, so lower-latency endpoints are chosen more often and higher temperatures explore more broadly. The EWMA is updated after every request, errors inflate the latency estimate, and endpoints with more than three consecutive failures receive near-zero probability. Unless otherwise noted, experiments use $\tau = 1.0$~ms and $\alpha = 0.3$.

The hypothesis is that this adaptive approach discovers the best endpoint by exploring and exploiting the latency signal. Section~\ref{sec:eval-boltzmann-pressure} shows where this hypothesis holds and where it does not. In a mixed read/write workload, many fast reads to a nearby follower can dominate the observed latency signal. The fewer writes sent to that same follower then do not move the average enough, even though they are still taking the wrong path. Boltzmann can therefore keep preferring a follower that looks good on reads while still being a poor write target. The same signal becomes useful for degradation detection (Section~\ref{sec:eval-partition}). When a node experiences packet loss or hardware failure, the latency increase is visible to the client and the router can use it to avoid that endpoint.

\subsection{Hybrid Routing Sends Writes to the Leader and Reads Across the Healthy Pool}
\label{sec:design-hybrid}

Hybrid is the policy implied by the operation-type asymmetry. It separates writes from reads.

\begin{itemize}
    \item \textbf{Writes.} Routed to the detected leader via the per-endpoint connection pool. This eliminates the forwarding hop entirely.
    \item \textbf{Reads.} Distributed across the current healthy read pool. In the normal case that means all members. If one member's read-side EWMA latency exceeds $\rho$ (default 2) times the cluster median EWMA, the router drops it from the read pool until it is no longer an outlier.
\end{itemize}

The read-side latency signal is therefore not asked to solve the steady-state placement problem. It acts only as a guardrail around the steady-state rule that reads should remain distributed. When all endpoints are healthy, hybrid reduces to the simple write-leader, read-distributed split. When one endpoint degrades, the same monitor prevents the read path from continuing to send it uniform traffic.

Leader detection and failover are active. A background poller issues a lightweight \texttt{Status()} across endpoints every 250~ms and refreshes the cached leader, and any leader-pinned write that fails clears the cached leader so writes fall back to the \texttt{round\_robin} shared client until the new leader is found. Section~\ref{sec:eval-leader-kill} shows why this matters. A client that cannot refresh the leader hint and fall back cleanly after an election is not a usable design, because startup-only detection leaves the write path pinned to a stale leader after leadership moves.

\section{Evaluation}
\label{sec:evaluation}

\subsection{Infrastructure}

Our refreshed main etcd comparisons use a 3-node etcd v3.5.17 cluster in us-west-2 with the same instance type and storage configuration on each node. The cluster spans multiple availability zones, and the client sits in the same region as the three members. The 5-node scale-up uses the same instance type and storage configuration. Smaller mechanism studies later in the section are called out explicitly when they are representative runs rather than fresh repeated-trial reruns.

\begin{table}[!t]
    \centering
    \footnotesize
    \caption{Experimental configuration.}
    \label{tab:cluster}
    \begin{tabular}{@{}lll@{}}
        \toprule
        \textbf{Node} & \textbf{Placement} & \textbf{Role} \\
        \midrule
        etcd-1 & same region & cluster member \\
        etcd-2 & same region & cluster member \\
        etcd-3 & same region & cluster member \\
        client & same region & benchmark runner \\
        \bottomrule
    \end{tabular}
\end{table}

\subsection{Workload}

Unless noted otherwise, each experiment runs for 60~seconds with 128 concurrent client workers. The default workload is 80\% linearizable point reads and 20\% single-key writes over a random 1{,}000-key keyspace. Each write stores a 256-byte value. We use 9 total gRPC connections on the 3-node cluster, 3 per endpoint for the distributed policies. On the 5-node cluster we keep the same per-endpoint budget, for 15 total connections. Connection counts are held equal across policies so routing, not connection count, explains the differences.

We use repeated trials for the paper's main quantitative comparisons. These cover the primary in-region comparison (Section~\ref{sec:eval-stats}), the balanced 50/50 rerun, the degraded-follower study (Section~\ref{sec:eval-partition}), the 5-node scale-up (Section~\ref{sec:eval-fivenode}), and the 5-node load sweep (Section~\ref{sec:eval-headroom}). In the primary healthy-cluster and degraded-follower measurements, the leader remained unchanged during the measured interval. We therefore read those tables as steady-state measurements of routing under a stable leader. Elections are handled separately in Section~\ref{sec:eval-leader-kill}. We retain representative runs for mechanism-oriented sweeps such as response size and leader-failover transients, where the effects are large relative to the run-to-run variance in the repeated-trial tables. The ZooKeeper section is supporting evidence rather than a precise estimate of small differences.

\begin{table}[t]
    \centering
    \footnotesize
    \caption{3-node steady-state means (5 trials per workload).}
    \label{tab:main}
    \begin{tabular}{@{}llrrr@{}}
        \toprule
        \textbf{Mix} & \textbf{Metric} & \textbf{round\_robin} & \textbf{hybrid} & \textbf{\% change} \\
        \midrule
        \multirow{3}{*}{80/20} & Read P50 (ms)  & $9.14$ & \textbf{$8.59$} & \textbf{$-6.0\%$} \\
                              & Write P50 (ms) & $6.22$ & \textbf{$4.39$} & \textbf{$-29.4\%$} \\
                              & Throughput (ops/s) & $14{,}536$ & \textbf{$15{,}809$} & \textbf{$+8.8\%$} \\
        \midrule
        \multirow{2}{*}{50/50} & Write P50 (ms) & $8.57$ & \textbf{$5.89$} & \textbf{$-31.3\%$} \\
                              & Throughput (ops/s) & $12{,}010$ & \textbf{$13{,}650$} & \textbf{$+13.7\%$} \\
        \bottomrule
    \end{tabular}
\end{table}

\subsection{Main In-Region Result}
\label{sec:eval-stats}

Table~\ref{tab:main} gives the refreshed in-region comparison on the current 3-node cluster, including both the main 80/20 workload and the balanced 50/50 rerun. The main comparison is stock \texttt{round\_robin} versus \texttt{hybrid}. In these runs the leader stayed fixed for the duration of the measured window, so the comparison isolates the steady-state cost of sending writes to followers first. Hybrid keeps the stock client's healthy read distribution, but corrects the write path. On the 80/20 workload, its write P50 is 4.39~ms, 29.4\% below \texttt{round\_robin}'s 6.22~ms, while read P50 stays close at 8.59~ms versus 9.14~ms. Throughput also rises from 14{,}536 to 15{,}809~ops/s, an 8.8\% gain.

The 50/50 rerun at the bottom of Table~\ref{tab:main} strengthens the same conclusion. Under that more write-heavy workload, hybrid lowers write P50 from 8.57~ms to 5.89~ms, a 31.3\% reduction relative to \texttt{round\_robin}, and raises throughput from 12{,}010 to 13{,}650~ops/s, a 13.7\% gain. Once writes are common enough to matter, the operation-type split becomes more valuable, not less.

\subsection{Boltzmann as a Supporting Comparator}
\label{sec:eval-boltzmann-pressure}

Boltzmann separates two claims. Latency adaptation helps once a read endpoint becomes visibly bad, but it is not a complete routing policy for mixed linearizable traffic because it has no operation-type distinction. A well-behaved latency-adaptive client still needs a write rule that follows the consensus protocol. The main comparison therefore centers on \texttt{round\_robin} versus \texttt{hybrid}.

\subsection{Read-Size Sweep Under Pure Reads}
\label{sec:eval-readsize}

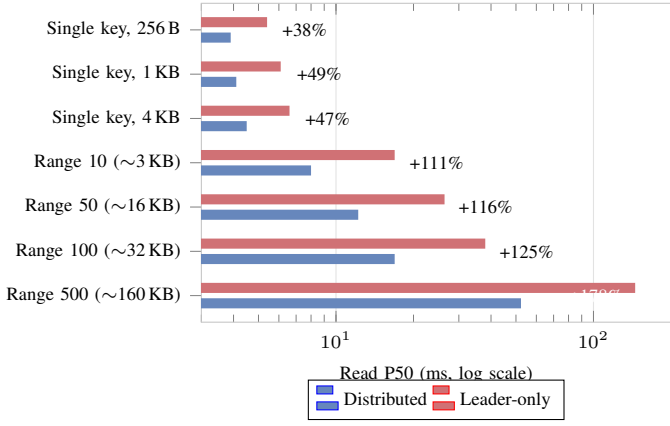
\begin{figure}[t]
  \centering
  \begin{tikzpicture}
    \begin{axis}[
      xbar,
      width=0.88\columnwidth,
      height=5.8cm,
      bar width=3.8pt,
      xmin=3,
      xmax=200,
      xmode=log,
      enlarge y limits=0.1,
      ytick={1,2,3,4,5,6,7},
      yticklabels={
        {Single key, 256\,B},
        {Single key, 1\,KB},
        {Single key, 4\,KB},
        {Range 10 ($\sim$3\,KB)},
        {Range 50 ($\sim$16\,KB)},
        {Range 100 ($\sim$32\,KB)},
        {Range 500 ($\sim$160\,KB)}
      },
      y dir=reverse,
      xlabel={Read P50 (ms, log scale)},
      tick label style={font=\scriptsize},
      label style={font=\scriptsize},
      legend style={
        font=\scriptsize,
        at={(0.5,-0.18)},
        anchor=north,
        legend columns=2
      },
      axis line style={draw=cMuted!50},
      major grid style={draw=cMuted!25},
      xmajorgrids=true,
    ]
      \addplot+[draw=none, fill=cRead!85] coordinates {
        (3.9,1)
        (4.1,2)
        (4.5,3)
        (8.0,4)
        (12.2,5)
        (16.9,6)
        (52.3,7)
      };
      \addlegendentry{Distributed}

      \addplot+[draw=none, fill=cLeader!80] coordinates {
        (5.4,1)
        (6.1,2)
        (6.6,3)
        (16.9,4)
        (26.4,5)
        (38.0,6)
        (145.2,7)
      };
      \addlegendentry{Leader-only}

      \node[font=\scriptsize, anchor=west] at (axis cs:5.7,1) {+38\%};
      \node[font=\scriptsize, anchor=west] at (axis cs:6.5,2) {+49\%};
      \node[font=\scriptsize, anchor=west] at (axis cs:7.0,3) {+47\%};
      \node[font=\scriptsize, anchor=west] at (axis cs:17.8,4) {+111\%};
      \node[font=\scriptsize, anchor=west] at (axis cs:27.6,5) {+116\%};
      \node[font=\scriptsize, anchor=west] at (axis cs:39.8,6) {+125\%};
      \node[font=\scriptsize, anchor=east, text=white] at (axis cs:142,7) {+178\%};
    \end{axis}
  \end{tikzpicture}
  \caption{Pure-read size sweep. Distributed linearizable reads beat leader-only
  reads at every response size, and the gap widens as larger responses stress
  the leader's local storage path.}
  \label{fig:readsize}
\end{figure}

To determine whether leader-direct or distributed routing is optimal for reads, we sweep response size with a 100\% read workload (0\% writes). We test both single-key reads at varying value sizes and range queries returning multiple keys. In Figure~\ref{fig:readsize}, ``single key'' means one \texttt{Get} returning a value of that size, while the range labels report the number of returned keys and the approximate total response payload.

The result is consistent across the sweep (Figure~\ref{fig:readsize}). We run one trial per size. Distributed reads win at \emph{every} response size, and the gap grows with response size: from 38\% at a 256-byte point read to 178\% at a 500-key range. The margins are much larger than the sub-millisecond run-to-run variance from Section~\ref{sec:eval-stats}, so the direction is unambiguous despite single runs. This matters because it separates the two halves of the routing rule. Write latency improves when writes stop being forwarded. Read serving improves when reads stop being concentrated.

\subsection{Recovery During Leader Election}
\label{sec:eval-leader-kill}

Leader-pinned write routing needs active leader re-detection after an election. In representative runs where we terminate the current Raft leader at T+30~s of a 90~s benchmark, a startup-only leader cache fails badly: writes keep timing out at the old leader endpoint and the client does not recover within the benchmark window. Hybrid therefore uses the failover from Section~\ref{sec:design-hybrid}, refreshing the cached leader with periodic \texttt{Status()} calls and falling back to the shared \texttt{round\_robin} client after a failed leader-pinned write. With that mechanism in place, \texttt{round\_robin} recovers in about 2~seconds with 23 in-flight errors, while hybrid recovers in about 3~seconds with 18 in-flight errors. We treat these as representative runs rather than a primary repeated-trial result, but they show that leader-pinned routing is usable only with active refresh and fallback, and that with those mechanisms its election behavior is close to stock \texttt{round\_robin}.

\subsection{Degraded Same-AZ Follower}
\label{sec:eval-partition}

To test how a nearby follower becoming slow changes the routing problem, we inject 8~ms of network delay (with 2~ms jitter) on etcd-1, a follower in the same AZ as both the client and the leader. This is a grey-failure case in which the node is still up but has become a bad read target. It is also the point where the stock client's healthy-cluster logic stops being sufficient, because continuing to send a third of reads to the slow follower is no longer the right choice. The leader again stayed unchanged during these measurements, so the result here reflects read-pool cleanup and steady write routing rather than election handling. We compare \texttt{round\_robin} and \texttt{hybrid} over 5 trials because the paper's main question here is whether the operation-aware split plus outlier ejection improves materially over stock routing.

\begin{table}[t]
    \centering
    \footnotesize
    \caption{Degraded-follower means (8 ms delay on etcd-1, 5 trials).}
    \label{tab:partition}
    \begin{tabular}{@{}lrrr@{}}
        \toprule
        \textbf{Metric} & \textbf{round\_robin} & \textbf{hybrid} & \textbf{\% change} \\
        \midrule
        Read P50 (ms)  & $8.62$ & \textbf{$8.19$} & \textbf{$-5.0\%$} \\
        Read P99 (ms)  & $47.02$ & \textbf{$16.84$} & \textbf{$-64.2\%$} \\
        Write P50 (ms) & $5.93$ & \textbf{$4.68$} & \textbf{$-21.1\%$} \\
        Write P99 (ms) & $34.31$ & \textbf{$8.97$} & \textbf{$-73.9\%$} \\
        Tput (ops/s)   & $8{,}377$ & \textbf{$15{,}841$} & \textbf{$+89.1\%$} \\
        Reads to delayed node & $\sim$33\% & \textbf{0.5\%} & \textbf{$\sim$-98\%} \\
        \bottomrule
    \end{tabular}
\end{table}

Table~\ref{tab:partition} summarizes the result over 5 trials. \texttt{round\_robin} continues to send about a third of traffic to the delayed same-AZ follower by construction, so its read P99 rises to 47.02~ms and its write P99 to 34.31~ms. Hybrid nearly eliminates reads to the slow member, only 0.5\% of aggregate read traffic, while preserving leader-pinned writes. That change cuts read P99 by 64.2\%, cuts write P50 by 21.1\%, cuts write P99 by 73.9\%, and raises throughput by 89.1\%. Read P50 stays close to the healthy-cluster level. The gain comes from avoiding a bad nearby endpoint and preserving the write path, not from making every healthy read faster.

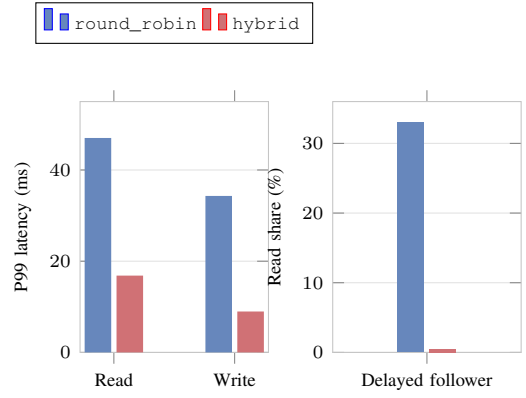
\begin{figure}[t]
  \centering
  \begin{tikzpicture}
    \begin{groupplot}[
      group style={group size=2 by 1, horizontal sep=0.85cm},
      width=0.46\columnwidth,
      height=4.9cm,
      ybar,
      tick label style={font=\scriptsize},
      label style={font=\scriptsize},
      axis line style={draw=cMuted!50},
      major grid style={draw=cMuted!25},
      ymajorgrids=true,
      enlarge x limits=0.28,
      legend style={
        font=\scriptsize,
        at={(0.5,1.23)},
        anchor=south,
        legend columns=2
      },
    ]
      \nextgroupplot[
        ymin=0,
        ymax=55,
        ylabel={P99 latency (ms)},
        symbolic x coords={Read,Write},
        xtick=data,
      ]
        \addplot+[draw=none, fill=cRead!85] coordinates {(Read,47.02) (Write,34.31)};
        \addplot+[draw=none, fill=cLeader!80] coordinates {(Read,16.84) (Write,8.97)};
        \legend{\texttt{round\_robin},\texttt{hybrid}}

      \nextgroupplot[
        ymin=0,
        ymax=36,
        ylabel={Read share (\%)},
        xmin=-0.6,
        xmax=0.6,
        xtick={0},
        xticklabels={Delayed follower},
      ]
        \addplot+[draw=none, fill=cRead!85] coordinates {(0,33.0)};
        \addplot+[draw=none, fill=cLeader!80] coordinates {(0,0.5)};
    \end{groupplot}
  \end{tikzpicture}
  \caption{Same-AZ degraded-follower result. The effect is not just lower
  median latency. Hybrid removes almost all reads from the slow follower and,
  as a result, sharply reduces both read and write tail latency. This is the
  paper's clearest operational gain.}
  \label{fig:partition}
\end{figure}

\subsection{Scaling to a 5-Node, 3-AZ Cluster}
\label{sec:eval-fivenode}

The experiments above use a 3-node cluster, the Raft minimum. To test whether the routing benefits hold or grow at scale, we deploy a 5-node cluster across three availability zones (etcd-1/etcd-3 in the client's AZ, etcd-2 cross-AZ, and two added members etcd-4/etcd-5 in a third AZ) and rerun the 80/20 workload over 5 trials. Connection budgets are held fair as before. The distributed policies use 3 connections per endpoint, 15 total.

\begin{table}[t]
    \centering
    \footnotesize
    \caption{5-node, 3-AZ means (80/20 read/write, 5 trials).}
    \label{tab:fivenode}
    \begin{tabular}{@{}lrrr@{}}
        \toprule
        \textbf{Metric} & \textbf{round\_robin} & \textbf{hybrid} & \textbf{\% change} \\
        \midrule
        Read P50 (ms)  & $9.76$ & \textbf{$9.41$} & \textbf{$-3.6\%$} \\
        Write P50 (ms) & $6.67$ & \textbf{$4.22$} & \textbf{$-36.7\%$} \\
        Tput (ops/s)   & $13{,}884$ & \textbf{$14{,}990$} & \textbf{$+8.0\%$} \\
        \bottomrule
    \end{tabular}
\end{table}

The write-side benefit remains large at 5 nodes (Table~\ref{tab:fivenode}). Over 5 trials, hybrid lowers write P50 by 36.7\%, from 6.67~ms to 4.22~ms, while improving throughput by 8.0\%. The leader remained fixed during these measurement windows as well. That result is what we should expect once the cluster grows from three members to five, because $4/5$ of \texttt{round\_robin}'s writes now land on a follower and pay the forwarding hop, versus $2/3$ at 3 nodes. Read P50 also stays close to stock distributed routing. The scale-up therefore strengthens the same systems point as the 3-node rerun, namely that the write-side forwarding cost does not disappear as the cluster grows.

\subsection{Load Sweep on 5 Nodes}
\label{sec:eval-headroom}

\begin{figure}[t]
  \centering
  \begin{tikzpicture}
    \begin{axis}[
      width=\columnwidth,
      height=5.4cm,
      xmin=110,
      xmax=530,
      ymin=3.8,
      ymax=10.2,
      xtick={128,256,512},
      xlabel={Concurrent workers},
      ylabel={Write P50 (ms)},
      tick label style={font=\scriptsize},
      label style={font=\scriptsize},
      legend style={
        font=\scriptsize,
        at={(0.5,-0.18)},
        anchor=north,
        legend columns=2
      },
      axis line style={draw=cMuted!50},
      major grid style={draw=cMuted!25},
      ymajorgrids=true,
    ]
      \addplot+[color=cRead!90, thick, mark=*, mark options={scale=0.9, fill=cRead!90}] coordinates {
        (128,6.72)
        (256,8.17)
        (512,9.70)
      };
      \addlegendentry{\texttt{round\_robin}}

      \addplot+[color=cLeader!90, thick, mark=square*, mark options={scale=0.8, fill=cLeader!90}] coordinates {
        (128,4.22)
        (256,5.33)
        (512,6.12)
      };
      \addlegendentry{\texttt{hybrid}}

      \node[font=\scriptsize, text=cLeader!90, anchor=west] at (axis cs:512,6.35) {3.58 ms gap};
    \end{axis}
  \end{tikzpicture}
  \caption{5-node load sweep. The write-latency gap does not close as
  concurrency rises. It widens. This is the clearest shape evidence that
  follower-forwarded writes consume avoidable leader capacity under load.}
  \label{fig:headroom}
\end{figure}
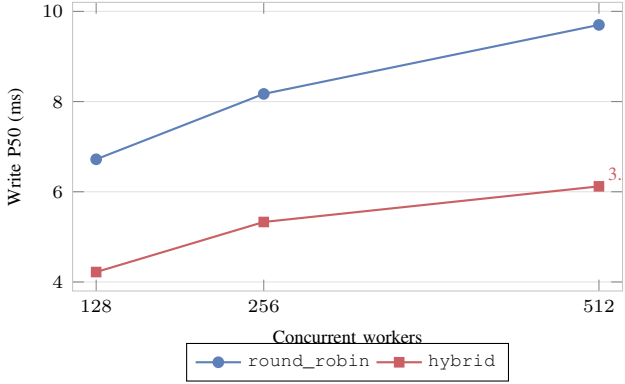

To test whether the mixed-workload write advantage persists as load rises, we reran the 5-node setup at 128, 256, and 512 workers for 30~seconds, again with 5 trials per point. This sweep is an independent rerun from Table~\ref{tab:fivenode}, so the 128-worker point is close to, but not identical to, the 128-worker values in that table. Figure~\ref{fig:headroom} is not a single-leader knee experiment, but it does provide a direct concurrency sweep on the real cluster.

The gap widens as concurrency rises. \texttt{round\_robin}'s write P50 climbs from 6.72~ms to 9.70~ms between 128 and 512 workers, while hybrid rises only from 4.22~ms to 6.12~ms. The absolute write-latency gap therefore grows from 2.50~ms to 3.58~ms.

\subsection{A Second System in Apache ZooKeeper}
\label{sec:eval-zookeeper}

The results so far are on etcd. To check that the operation-type asymmetry is not peculiar to one codebase, we ported the three paper policies to Apache ZooKeeper~3.9.2 (ZAB) and repeated the primary experiment on a fresh implementation. The semantic mapping is direct (Section~\ref{sec:bg-zk}). A write forwards to the ZAB leader as a Raft proposal does. A linearizable read is a \texttt{sync()} followed by a local \texttt{get()}, with \texttt{sync()} serving as the ReadIndex analogue. We ran a fresh 3-node ensemble with a dedicated client in a multi-AZ region, 80/20 read/write, 128 workers, over 3 trials and use it as supporting cross-system evidence rather than as a primary statistical result. In what follows, this implementation is the experimental ZooKeeper client.

\begin{table}[t]
    \centering
    \footnotesize
    \caption{ZooKeeper 3-node results (80/20 read/write, 3 trials).}
    \label{tab:zookeeper}
    \begin{tabular}{@{}lrrr@{}}
        \toprule
        \textbf{Mode} & \textbf{Write P50 (ms)} & \textbf{Read P50 (ms)} & \textbf{Tput (ops/s)} \\
        \midrule
        round\_robin  & $3.94 \pm 0.06$ & $6.89 \pm 0.12$ & $21{,}043 \pm 756$ \\
        boltzmann     & $4.14 \pm 0.02$ & $7.53 \pm 0.15$ & $18{,}574 \pm 361$ \\
        hybrid        & $\mathbf{3.67 \pm 0.01}$ & $\mathbf{6.81 \pm 0.02}$ & $20{,}393 \pm 69$ \\
        \bottomrule
    \end{tabular}
\end{table}

The pattern reproduces (Table~\ref{tab:zookeeper}). In the 3-trial ZooKeeper run, hybrid again has the lowest write latency. Its write P50 is 3.67~ms, 6.9\% below \texttt{round\_robin}'s 3.94~ms, while the latency-adaptive policy again trails on writes. We use this result for direction rather than for fine-grained statistical separation.

That table still uses the experimental ZooKeeper client rather than ZooKeeper's literal upstream client shape. The upstream client opens one session to one server from the connect string and stays there until reconnect, so we also checked a sticky session pinned to a confirmed same-AZ follower. In the healthy 80/20 rerun over 3 trials, that sticky session averaged 12.10~ms read P50 and 6.32~ms write P50, versus 7.13~ms and 3.97~ms for hybrid, with throughput 11.68k versus 19.19k~ops/s. The stronger result came when that follower degraded, because the upstream client kept the entire session on the now-slow member while hybrid could shift reads away from it and keep writes on the leader.

\begin{table}[t]
    \centering
    \footnotesize
    \caption{ZooKeeper degraded-follower results (3 trials).}
    \label{tab:zookeeper-sticky}
    \begin{tabular}{@{}lrrr@{}}
        \toprule
        \textbf{Mode} & \textbf{Read P50 (ms)} & \textbf{Write P50 (ms)} & \textbf{Tput (ops/s)} \\
        \midrule
        sticky follower & $45.95 \pm 0.54$ & $23.36 \pm 0.23$ & $3{,}075 \pm 30$ \\
        hybrid          & $\mathbf{8.01 \pm 0.06}$ & $\mathbf{4.50 \pm 0.01}$ & $\mathbf{16{,}915 \pm 77}$ \\
        \bottomrule
    \end{tabular}
\end{table}

Table~\ref{tab:zookeeper-sticky} turns that sticky-session check into a repeated-trial result. With 8~ms of delay and 2~ms of jitter on the pinned same-AZ follower, the sticky client rises to 45.95~ms read P50 and 23.36~ms write P50, while hybrid stays at 8.01~ms and 4.50~ms. Throughput rises from 3.08k to 16.91k~ops/s. Relative to the sticky session, hybrid lowers read P50 by 82.6\%, lowers write P50 by 80.7\%, and raises throughput by 5.5$\times$.

\subsection{End-to-End Through kube-apiserver}
\label{sec:eval-apiserver}

\begin{table}[t]
    \centering
    \footnotesize
    \caption{Standalone kube-apiserver results (watch cache disabled, GET/POST).}
    \label{tab:apiserver}
    \begin{tabular}{@{}llrrr@{}}
        \toprule
        \textbf{Mix} & \textbf{Mode} & \textbf{Read P50} & \textbf{Write P50} & \textbf{Tput} \\
        \midrule
        80/20 r/w & stock  & 8.26~ms & 6.95~ms & 7{,}588 \\
        80/20 r/w & hybrid & 8.70~ms & \textbf{6.37~ms} & 7{,}196 \\
        20/80 r/w & stock  & 13.18~ms & 10.18~ms & 5{,}788 \\
        20/80 r/w & hybrid & 13.76~ms & \textbf{8.61~ms} & \textbf{5{,}876} \\
        \bottomrule
    \end{tabular}
\end{table}

To test whether the routing effect survives a real control-plane hop, we patch kube-apiserver's etcd storage client with the same hybrid split and drive it with a standalone REST load generator. We disable the watch cache and issue only GET and POST operations, no LIST requests, so both reads and writes stay on the uncached etcd path. We compare the same patched binary with hybrid routing disabled (\texttt{stock}) and enabled (\texttt{hybrid}). Table~\ref{tab:apiserver} reports representative 60-second runs at 64-way concurrency. The end-to-end result is narrower than the direct-client result but still consistent with it. In the 80/20 mix, hybrid lowers write P50 from 6.95~ms to 6.37~ms, an 8.3\% reduction, while slightly reducing throughput and slightly increasing read latency. In the 20/80 mix, where the workload exercises the write path more heavily, hybrid lowers write P50 from 10.18~ms to 8.61~ms, a 15.4\% reduction, and throughput rises slightly from 5{,}788 to 5{,}876~ops/s. We do not read this as evidence that kube-apiserver becomes uniformly faster under hybrid routing. Instead, it shows that the write-forwarding penalty remains visible end to end when etcd stays on the critical path, while apiserver-side work and uncached read handling dilute the gain in read-heavier mixes. We therefore use this section as supporting evidence, not as the paper's headline result.

\section{Discussion}
\label{sec:discussion}

\subsection{Why the Split Works}

The read-size sweep (Section~\ref{sec:eval-readsize}) explains the two parts of the rule. ReadIndex is a lightweight confirmation rather than a forwarding of the query, so distributing linearizable reads spreads the real read-serving work across the cluster. Writes behave differently because a follower only forwards them to the leader. The degraded-follower case adds a third point. Once one member becomes a bad read target, the client should stop treating it as an equal read destination without changing the write rule.

\subsection{What the Latency Signal Should Do}

Client-observed latency is not a complete steady-state routing signal for mixed linearizable traffic. Many fast reads to a nearby follower can dominate the observed signal even while the fewer writes sent there still pay forwarding through the leader. That is why the paper centers on \texttt{round\_robin} versus \texttt{hybrid}. The same signal is still useful for degradation detection. With 8~ms of induced delay on a same-AZ follower, hybrid cuts reads to that member from the uniform one-third share to 0.5\% and sharply improves both read and write tails.

\subsection{Scope}

The paper makes two claims, and it is better to keep them separate. The steady-state claim is that when the leader stays fixed, follower forwarding is wasted work on the write path while distributed reads still make sense. That is what the main 3-node, 5-node, and degraded-follower tables measure. The second claim is about usability under leader change. Section~\ref{sec:eval-leader-kill} shows that a startup-only leader cache leaves the write path stalled after leader loss, while the hybrid failover path recovers within about 3~seconds and only tens of in-flight errors, close to \texttt{round\_robin}. We validate the result primarily on 3-node and 5-node etcd clusters, with supporting checks in ZooKeeper. We do not claim a full scaling law, and some mechanism studies remain representative runs rather than repeated-trial means.

\subsection{Practical Impact: Follow-on etcd Design}

The routing principle evaluated in this paper has since informed a concrete
follow-on design discussion in the etcd project. Open issue~\#22268 proposes a
two-phase implementation: first, expose an advisory \texttt{leader\_id} in the
v3 \texttt{ResponseHeader}; second, add an opt-in client balancer that routes
leader-dependent operations to that member while leaving reads, watches, and
keep-alives distributed~\cite{etcd-issue-22268}. Stale, failed, or ambiguous
hints fall back to round-robin, preserving existing retry, forwarding, Raft,
and watch behavior.

The follow-on prototype reports 24.9\% fewer peer bytes for 64-KiB
\texttt{Put} requests and 101/101 successful writes with a paused follower,
compared with 67/101 under round-robin. These results provide a concrete
engineering path from the paper's protocol-aware routing principle to an
upstream client design.

\section{Related Work}
\label{sec:related}

\textbf{etcd client routing.} etcd issue \#15918~\cite{etcdissue15918} showed that round-robin causes 66\% cross-zone traffic and proposed ``prefer local member.'' Issue \#7109~\cite{etcdissue7109} measured that 78\% of linearizable read latency is the ReadIndex wait. PR \#21765 showed that connection pooling (2 connections) improved large list operations from minutes to seconds. We do not claim that leader-directed writes or follower-served reads are individually new. What is missing is a client-side rule that separates reads from writes and measures where that split helps.

\textbf{Consensus and read scalability.} Leader-based consensus descends from Paxos~\cite{lamport1998parttime} and the state-machine replication model~\cite{schneider1990statemachine}, realized in Raft~\cite{etcdraft}, Zab~\cite{junqueira2011zab}, and the Chubby lock service~\cite{burrows2006chubby}. Ongaro's PhD thesis~\cite{ongaro2014consensus} (Section 6.4) proposes follower-served reads to offload the leader, which etcd implements via ReadIndex. Leaderless and multi-leader protocols such as EPaxos~\cite{moraru2013epaxos} spread the write load itself, and later work extends Raft to serve reads from a majority rather than the leader alone~\cite{arora2017leaderless}. Both change the protocol, whereas our work is a client-side routing rule that leaves the protocol untouched. Because a follower can serve a linearizable read~\cite{herlihy1990linearizability} after the ReadIndex confirmation, keeping reads distributed across the healthy read pool appears to be the right in-region policy across the response sizes we measure, and the benefit grows with size as the leader's serving cost dominates the confirmation.

\textbf{Read-optimized replication.} Chain replication~\cite{vanrenesse2004chainreplication} and CRAQ~\cite{terrace2009objectstore} restructure replication so that reads scale across replicas while writes serialize through a designated node, a read/write asymmetry in spirit similar to ours but achieved by changing the replication topology rather than by client routing over an unchanged consensus cluster.

\textbf{Kubernetes optimizations.} Consistent Reads from Cache (KEP-2340)~\cite{kep2340}, WatchList (KEP-3157), and Snapshottable API Server Cache (KEP-4988) reduce etcd read load at the apiserver level. These are orthogonal. They reduce the number of requests that reach etcd, while we optimize routing for the requests that remain.

\textbf{Adaptive load balancing.} Power-of-two-choices~\cite{mitzenmacher2001power}, weighted-round-robin with server-reported load, and latency-based routing (EWMA, peak-EWMA) are standard in stateless systems. Finagle~\cite{finagle} uses latency-weighted P2C for RPC services. These assume the cheapest endpoint from the client's perspective is the cheapest end-to-end, which consensus-cost invisibility shows does not hold for linearizable operations. Our results show where the latency signal does belong. It is useful for degradation detection, where it provides graduated circuit-breaker behavior that improves on binary thresholds.

\textbf{Topology-aware routing in service meshes.} Istio locality load balancing and Kubernetes topology-aware hints (KEP-2433) route by AZ proximity. These reduce cross-AZ traffic but remain operation-blind. They treat reads and writes alike and ignore leader placement. A topology-aware policy that prefers the same-AZ follower would still pay the ReadIndex RTT to a cross-AZ leader.

\textbf{Range-based and read-mostly systems.} CockroachDB~\cite{taft2020cockroachdb} follower reads~\cite{cockroachfollowerreads} and TiKV's learner reads trade consistency for lower read latency, while our approach keeps linearizable reads and writes on their existing consensus path. Read-mostly caching tiers such as Facebook's memcache deployment~\cite{nishtala2013memcache} attack read load at a different layer, complementary to routing within the store.

\section{Conclusion}
\label{sec:conclusion}

Reads and writes place opposite demands on a consensus leader, and a client should route accordingly. A write commits through the leader, so routing it to a follower only adds a forwarding hop. A linearizable read needs only a lightweight confirmation from the leader, after which any healthy member can serve it from local state. In the steady state, linearizable reads should stay distributed while writes should go to the leader. Once a member becomes clearly degraded, it should leave the read pool.

Latency-only routing fails because the blended client-observed signal does not separate read-serving cost from write-forwarding cost, though it remains useful for degradation detection.

Our strongest result is the degraded same-AZ follower case. Over 5 trials, \texttt{hybrid} nearly stops reading from the slow member and cuts read P99 by 64.2\%, cuts write P99 by 73.9\%, and raises throughput by 89.1\% relative to stock \texttt{round\_robin}. In the healthy in-region case, hybrid lowers write P50 by 29.4\% on 3 nodes and 36.7\% on 5 nodes while keeping distributed linearizable reads. Those steady-state measurements were taken while the leader stayed fixed, so we read them as evidence about forwarding cost, not as evidence that startup-only leader tracking is sufficient. Once leadership changes, the client still needs active refresh and fallback. The supporting ZooKeeper results point in the same direction, suggesting that the routing rule follows from leader-based consensus structure rather than one implementation. In a replicated store, client routing should follow the protocol work an operation induces, not only the latency the client happens to observe.

\bibliographystyle{IEEEtran}
{\footnotesize \bibliography{references}}

\end{document}